\documentstyle[epsfig]{aipproc}

\begin{document} 
\title{The Spectral Variability of\\Cygnus X-1 at MeV Energies}

\author{M.L. McConnell$^*$, K. Bennett$^{**}$, H. Bloemen$^{\dagger}$, 
W. Collmar$^{\dagger\dagger}$, W. Hermsen$^{\dagger}$, 
L. Kuiper$^{\dagger}$,  
B. Phlips$^{\ddag}$, J.M. Ryan$^*$, V. Sch\"onfelder$^{\dagger\dagger}$, 
\\H. Steinle$^{\dagger\dagger}$, A.W. Strong$^{\dagger\dagger}$}

\address{$^*$Space Science Center, University of New Hampshire, Durham, NH  03824, USA\\ 
$^{**}$Space Science Department, ESTEC, Noordwijk, The Netherlands\\
$^{\dagger}$Space Research Organization of the Netherlands (SRON), Utrecht, The Netherlands\\
$^{\dagger\dagger}$Max Planck Institute for Extraterrestrial Physics (MPE), Garching, Germany\\
$^{\ddag}$George Mason University, Fairfax, VA  22030, USA}

%\lefthead{LEFT head} 
%\righthead{RIGHT head} 
\maketitle

\begin{abstract} 
In previous work, we have used data from the first three years of the 
CGRO mission to assemble a broad-band $\gamma$-ray spectrum of the 
galactic black hole candidate Cygnus X-1.  Contemporaneous data 
from the COMPTEL, OSSE and BATSE experiments on CGRO were  
selected on the basis of the hard X-ray flux (45--140 keV) 
as measured by BATSE.  This provided a spectrum of Cygnus X-1 in 
its canonical low X-ray state (as measured at energies below 10 keV), covering the energy range 
from 50 keV to 5 MeV.   Here we report on a comparison of this 
spectrum to a COMPTEL-OSSE spectrum collected during a high X-ray state of
Cygnus X-1 (May, 1996).  These data provide evidence for significant 
spectral variability at energies above 1 MeV.  
In particular, whereas the hard X-ray flux {\it decreases} during 
the high X-ray state, the flux at energies above 1 MeV {\it increases}, 
resulting in a significantly harder high energy spectrum. 
This behavior is consistent with the general picture of galactic black hole
candidates having two distinct spectral forms at soft $\gamma$-ray energies.
These data extend this picture, for the first time, to energies 
above 1 MeV. 
\end{abstract}

\section*{Introduction}

Observations by the instruments on CGRO, coupled with observations  
by other high-energy experiments (e.g., SIGMA, ASCA and RXTE)
have provided a wealth of new information 
regarding the emission properties of galactic black hole candidates.  
An important aspect of these high energy radiations is spectral variability,
observations of which can provide constraints on models which seek to describe
the global emission processes.  Based on observations 
by OSSE of seven transient galactic 
black hole candidates at soft $\gamma$-ray energies (i.e., below 1 MeV), 
two $\gamma$-ray spectral shapes have been identified that appear 
to be well-correlated with the soft X-ray state \cite{Grove1997,Grove1998}.  
In particular, these observations define a {\it breaking}
$\gamma$-ray spectrum that corresponds to the low X-ray state and a 
{\it power-law} $\gamma$-ray spectrum that corresponds to the high X-ray state.
(Here we emphasize that the 'state' is that measured at soft X-ray energies, below 10 keV.)

At X-ray energies, the measured flux from Cyg X-1 is known to be variable over a 
wide range of time scales,
ranging from msec to months.  It spends most of its time in a low X-ray state, 
exhibiting a breaking spectrum at $\gamma$-ray energies that is often characterized as 
a Comptonization spectrum.
In May of 1996, a transition of Cyg X-1 into a high X-ray state was observed by RXTE, beginning 
on May 10 \cite{Cui1997}.  The 2--12 keV flux reached a level of 2 Crab on May 19, 
four times higher than its normal value. Meanwhile, at hard X-ray energies (20-200 keV), 
BATSE measured a
significant {\it decrease} in flux \cite{Zhang1997}.  Motivated by these dramatic changes, a 
target-of-opportunity (ToO) for CGRO, with observations by OSSE, COMPTEL and EGRET,
began on June 14 (CGRO viewing period 522.5). 
Here we report on the results from an analysis of the COMPTEL data from this ToO 
observation.

\section*{Observations and Data Analysis}

COMPTEL has obtained numerous observations of the Cygnus region since its launch in 1991,
providing the best available source of data for studies of Cyg X-1 at energies above 1 
MeV.  Figure 1 
shows a plot of hard X-ray flux, as obtained from BATSE occultation monitoring, for 
each day in which Cyg X-1 was within 40$^{\circ}$ of the COMPTEL pointing direction.

\begin{figure}
\centerline{\epsfig{file=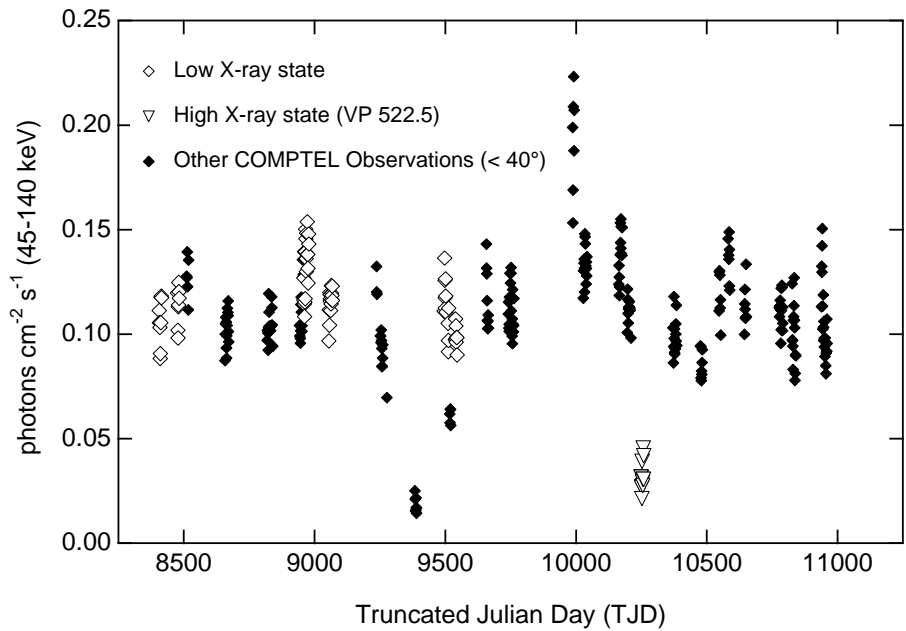,width=5.0in}}
\vspace{0.1in}
\caption{Hard X-ray time history (from 45--140 keV BATSE occultation data) for COMPTEL observations of Cyg X-1.
Open diamonds indicate those data used to generate the low-state $\gamma$-ray spectrum. 
Open triangles correspond to CGRO viewing period 522.5.}
\end{figure}

In previous work, we have compiled
a broad-band spectrum of Cyg X-1 using contemporaneous data from BATSE, OSSE and COMPTEL
\cite{McConnell1999,McConnell2000}. The observations were chosen, in part, based on the 
level of hard X-ray flux measured by BATSE, the goal being to ensure a spectral measurement 
that corresponded to a common spectral state.  In Figure 1, the data points from the
selected observations are indicated by open diamonds.
The resulting spectrum, corresponding to a low X-ray state, showed evidence for 
emission out to 5 MeV.  The spectral shape, although consistent with the so-called breaking 
spectral state \cite{Grove1997,Grove1998}, was clearly not consistent with standard 
Comptonization models. The COMPTEL data provided evidence for a hard tail at energies 
above $\sim$1 MeV that extended to perhaps 5 MeV.

During the high X-ray state observations in May of 1996 (VP 522.5), COMPTEL collected 11 days of 
data at a favorable aspect angle of 5.3$^{\circ}$.  The hard X-ray flux for these days is 
denoted by open triangles in Figure 1.
An analysis of COMPTEL data from this observation revealed some unusual characteristics.  The 
1--3 MeV image (Figure 2) showed an unusually strong signal from Cyg X-1 when 
compared with other observations of similar exposure.  The flux level was 
significantly higher than the average flux seen from earlier observations
\cite{McConnell1999,McConnell2000}.  In the 1--3 MeV energy 
band, the flux had increased by a factor of 2.5, from $8.6 (\pm2.7) \times 10^{-5}$ cm$^{-2}$
s$^{-1}$ MeV$^{-1}$ to $2.2 (\pm0.4) \times 10^{-4}$ cm$^{-2}$ s$^{-1}$ MeV$^{-1}$.  The 
observed change in flux is significant at a level of $2.6\sigma$.
In addition, unlike in previous measurements, there was no evidence for any emission at 
energies {\it below} 1 MeV.
This fact is explained, in part,  by a slowly degrading sensitivity of 
COMPTEL at energies below 1 MeV due to increasing energy thresholds in the 
lower (D2) detection plane.  
Part of the explanation, however, appears to be a much harder 
source spectrum.

\begin{figure}
\epsfig{file=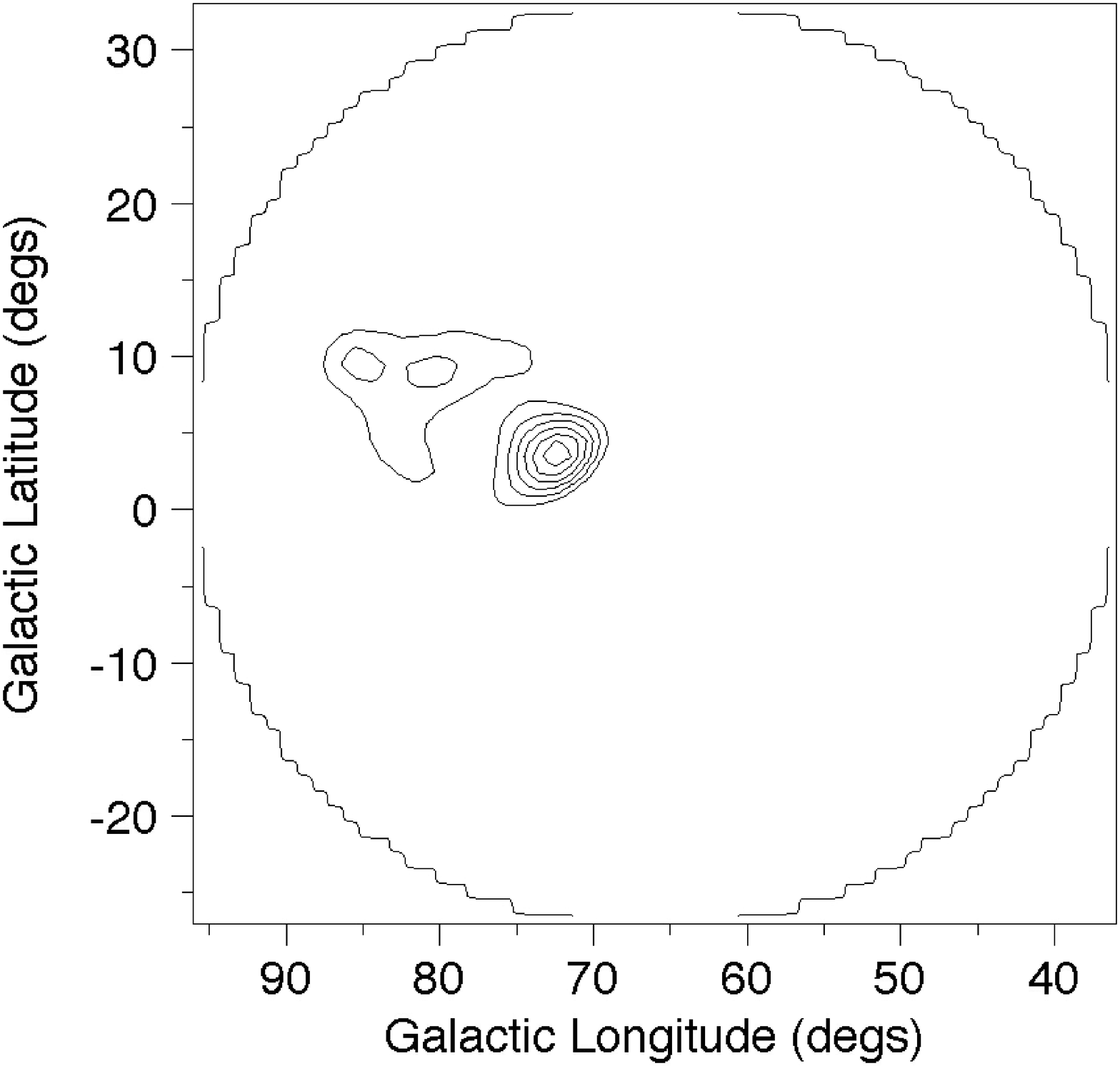,width=2.6in}
\hfill
\epsfig{file=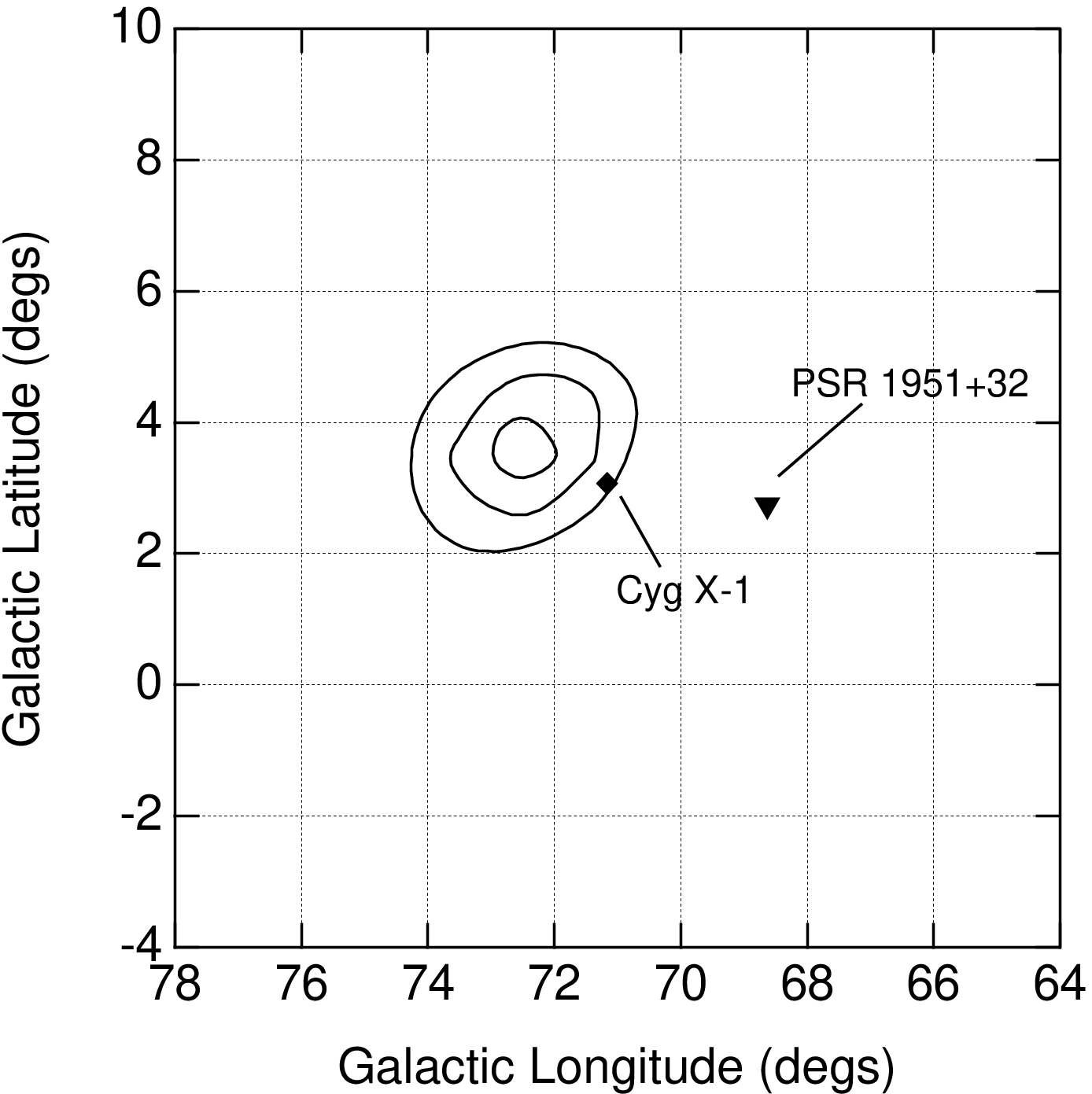,width=2.6in}
\vspace{0.2in}
\caption{COMPTEL imaging of the Cygnus region as derived from 1--3 MeV data collected during
high X-ray state of VP 522.5.  The left-hand figure shows the maximum likelihood map. 
The right-hand figure shows the 1, 2 and 3-$\sigma$ location contours.  The emission is consistent
with a point source at the location of Cyg X-1, with no significant contribution from PSR 1951+32.}
\end{figure}

A more complete picture of the MeV spectrum is obtained by combining the 
COMPTEL results with results from OSSE, extending the measured spectrum 
down to $\sim$50 keV.  Unfortunately, a comparison of the COMPTEL and OSSE spectra 
for VP 522.5 shows indications for an 
offset between the two spectra by about a factor of two, with the OSSE flux points being lower 
than those of COMPTEL in the overlapping energy region near 1 MeV.  A similar offset between OSSE 
and COMPTEL-BATSE is also evident in the contemporaneous low soft X-ray state spectrum 
\cite{McConnell1999,McConnell2000}.  The origin of this offset is not clear.  Here we shall 
assume that there exists some uncertainty in the instrument calibrations and that this 
uncertainty manifests itself in a global normalization offset.  We have subsequently
increased the flux for each OSSE data point by a factor of two.   This provides a good 
match between COMPTEL and OSSE at 1 MeV for both the low-state and high-state spectra,
but we are left with an uncertainty (by a factor of 
two) in the absolute normalization of the spectra.

We compare the resulting COMPTEL-OSSE spectra in Figure 3 (with the data points in both 
OSSE spectra increased by a factor two).  The low-state spectrum 
shows the breaking type spectrum that is typical of most high energy observations 
of Cyg X-1.  The high-state spectrum, on the other hand, shows the power-law type 
spectrum that is characteristic of black hole candidates in their high X-ray 
state.  This spectral behavior had already been reported for this time period 
based on observations with both BATSE \cite{Zhang1997b} and OSSE\cite{Gierlinski1997}.
The inclusion of the COMPTEL data 
provides  evidence, for the first time, of a continuous power-law 
(with a photon spectral index of -2.6) extending beyond 1 MeV, 
up to $\sim$10 MeV. 

A  power-law spectrum had also been observed by both OSSE and BATSE
during February of 1994 \cite{Phlips1996,Ling1997}, corresponding to the low level of 
hard X-ray flux near TJD 9400 in Figure 1. In this case, however, the amplitude of the 
power-law was too low for it to be detected by COMPTEL.

\begin{figure}
\centerline{\epsfig{file=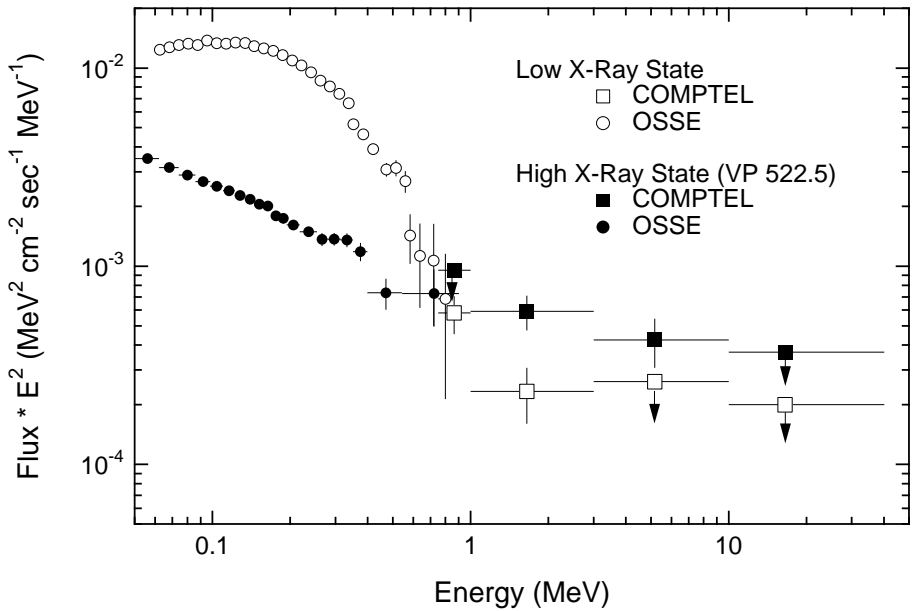,width=5.0in}}
\caption{Spectra of Cyg X-1, shown as $E^2$ times the photon flux.  OSSE flux levels 
have been increased by a factor of two and OSSE upper 
limits have been removed for the sake of clarity.}
\end{figure}

\section*{Discussion}

We can use the COMPTEL data alone to draw some important conclusions regarding the 
MeV variability of Cyg X-1.  Most importantly, the flux measured by COMPTEL at 
energies above 1 MeV
was observed to be higher (by a factor of 2.5) during the high X-ray state 
(in May of 1996) than it was during the low X-ray state.  The lack of any detectable
emission below 1 MeV further suggests a relatively hard spectrum.

Inclusion of the OSSE spectra clearly show an evolution from a 
breaking type spectrum in the low X-ray state to a power-law spectrum in 
the high X-ray state.  The COMPTEL data are 
consistent  with a pivot point near 1 MeV.  The 
power-law appears to extend to $\sim$10 MeV with no clear indication of 
a cut-off.

\section*{Acknowledgements}

The COMPTEL project is supported by NASA under contract NAS5-26645, by the
Deutsche Agentur f\"ur Raumfahrtgelenheiten (DARA) under grant 50 QV90968 and by
the Netherlands Organization for Scientific Research NWO.  This work was also 
supported by NASA grant NAG5-7745.

\end{document}